\documentclass[aps,prl,reprint,superscriptaddress,longbibliography,nobibnotes,nofootinbib]{revtex4-2}
\usepackage{graphicx}
\usepackage{amsmath}
\usepackage[hidelinks]{hyperref}
\usepackage{float}
\makeatletter
\newif\ifendmattertitle

\def\maketitle{%
  \@author@finish
  \ifendmattertitle
    \begingroup
      \let\@AAC@list\@empty
      \let\@AFF@list\@empty
      \let\@AFG@list\@empty
      \let\@FMN@list\@empty

      \let\frontmatter@RRAPformat\@gobble

      \title@column\titleblock@produce
    \endgroup
    \global\endmattertitlefalse
  \else
    \title@column\titleblock@produce
  \fi
  \suppressfloats[t]
}

\makeatother

\begin{document}

\title{A Unified Subject Map for 130 Years of Physics}

\author{Khoa Nguyen}
\thanks{These authors contributed equally to this work.}
\affiliation{Department of Electrical Engineering and Computer Sciences, University of California, Berkeley, California 94720, USA}

\author{Pragyan Pandey}
\thanks{These authors contributed equally to this work.}
\affiliation{Department of Physics, University of California, Berkeley, California 94720, USA}

\author{Sophie Li}
\affiliation{Palo Alto High School, Palo Alto, California 94301, USA}

\author{Eric Y. Ma}
\email{Contact author: eric.y.ma@berkeley.edu}
\affiliation{Department of Electrical Engineering and Computer Sciences, University of California, Berkeley, California 94720, USA}
\affiliation{Department of Physics, University of California, Berkeley, California 94720, USA}
\affiliation{Lawrence Berkeley National Laboratory, Berkeley, California 94720, USA}

\date{\today}
\begin{abstract}
More than a century of physics is recorded in the American Physical Society (APS) archive, but the corpus cannot be analyzed as a single, time-resolved object because its subject metadata are fragmented across eras with no shared vocabulary. We close this gap by using a frontier large language model to retrospectively assign the modern Physics Subject Headings (PhySH) to the historical archive, yielding a unified subject map for every APS paper from 1893 to 2025. The resulting map not only reproduces century-scale disciplinary arcs but also resolves the fine-grained lifecycles of individual ideas, materials, techniques, and discoveries across a vocabulary of over 3,000 PhySH Concepts. The map turns a fragmented archive into a quantitative substrate for systematic search and for data-driven studies of how physics evolves.
\end{abstract}

\maketitle

\textit{Introduction---}The American Physical Society (APS) archive is one of the most continuous records of modern physics. Its more than 750,000 peer-reviewed papers, published from 1893 to the present \cite{APSArchivePolicy}, chronicle the growth of the field for more than a century. Yet the archive is not naturally a single analyzable object. To compare papers across eras, one needs subject metadata expressed in a shared vocabulary, but most APS papers lack this.

Since 2016, APS has organized $\sim$200,000 new papers using Physics Subject Headings (PhySH), a faceted taxonomy designed for modern physics that assigns papers fine-grained, hierarchically organized \textit{concepts} \cite{physh_about,physh_home,APSNews2016PhySH,PRL2016EditorialReminders}. A paper can be labeled by multiple concepts, which are grouped under two dimensions: \textit{facets}, which distinguish the role of a concept in the paper, such as Research Areas, Physical Systems, and Properties; and \textit{disciplines}, which place the work within broader fields of physics such as Condensed Matter Physics and Particles \& Fields [Fig.~\ref{fig:physh_and_gap}(a,b)].

The remaining $\sim$550,000 pre-2016 papers, however, were never annotated under PhySH [Fig.~\ref{fig:physh_and_gap}(c)]. Many carry codes from the deprecated Physics and Astronomy Classification Scheme (PACS), which APS retired mainly because it failed to keep pace with the emergence of new subfields \cite{APSNews2016PhySH,PRL2016EditorialReminders}; the earliest era of the archive carried no subject metadata at all. As a result, the historical corpus cannot be systematically searched, compared, or aggregated under a single subject vocabulary, and questions about how individual ideas, materials, and techniques have moved through physics over century-long timescales are difficult to ask quantitatively \cite{Redner2005CitationStats,Herrera2010MappingFields,Pan2012Interdisciplinarity,Sinatra2015CenturyPhysics,Battiston2019CensusPhysics}.

\begin{figure}[H]
    \centering
    \includegraphics[width=\linewidth]{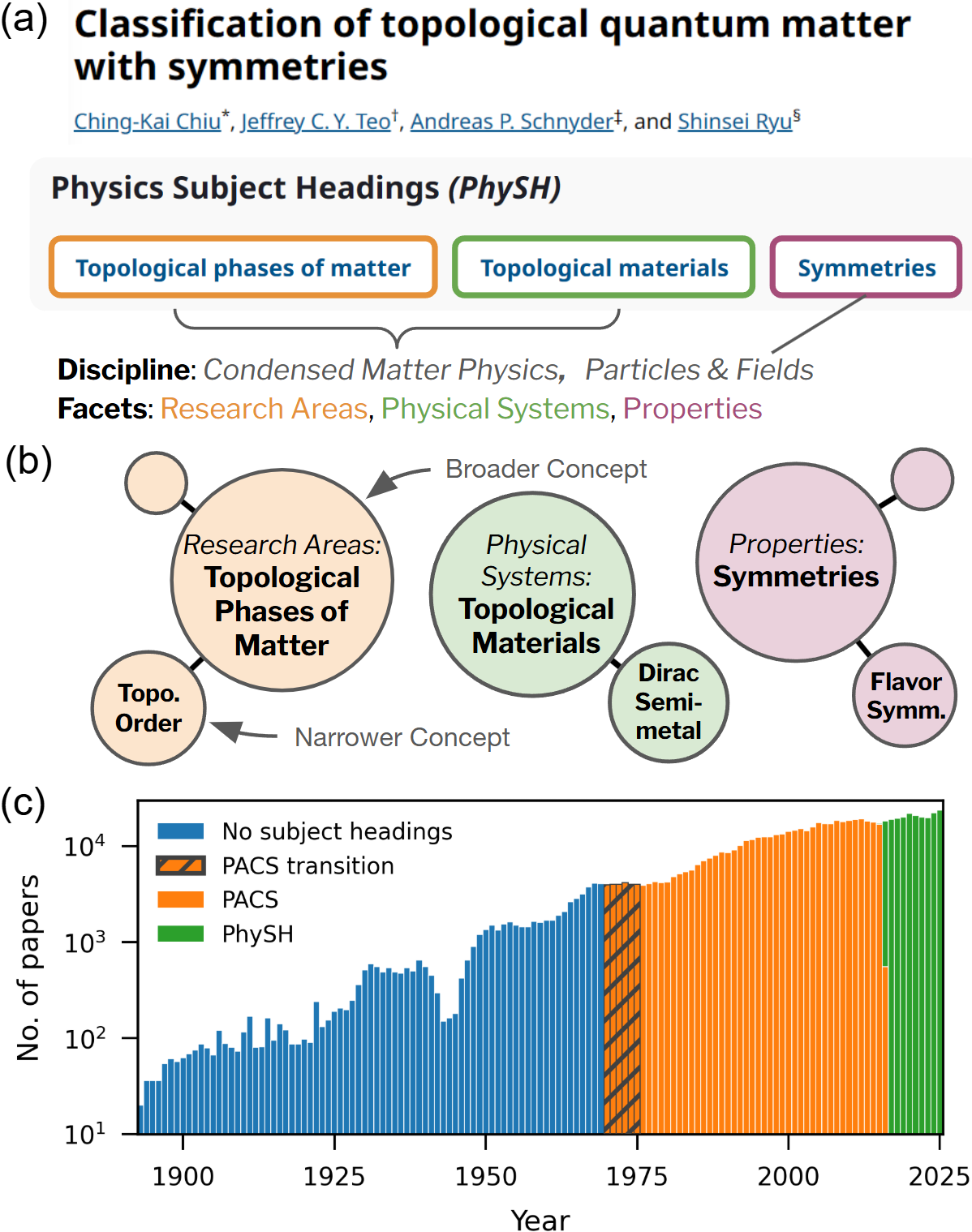}
    \caption{\textbf{PhySH scheme and the APS metadata gap.}
    (a) Example APS paper annotated using PhySH \cite{Chiu2016TopologicalQuantumMatter}.
    (b) PhySH concepts are hierarchical and organized by facets.
    (c) APS papers by year, colored by available subject metadata.}
    \label{fig:physh_and_gap}
\end{figure}

In this Letter, we close this gap by retrospectively applying PhySH to every APS paper since 1893 using a frontier large language model (LLM). The resulting map is, to our knowledge, the first continuous, fine-grained subject classification of the APS archive over its full 130-year span. 

\textit{Methods---}APS distributes titles, abstracts, and bibliographic metadata for all 756{,}183 papers in its journals from 1893 to 2025 \cite{aps_corpus}. Of these, 205,498 post-2016 papers carry author-assigned PhySH labels while the remaining 550,685 require retrospective labeling. 

While bespoke supervised classifiers could plausibly label the 17 disciplines, multi-label assignment across 3{,}000+ PhySH concepts is where the semantic prior of a pretrained LLM is especially valuable.

\begin{figure}[t]
    \centering
    \includegraphics[width=\linewidth]{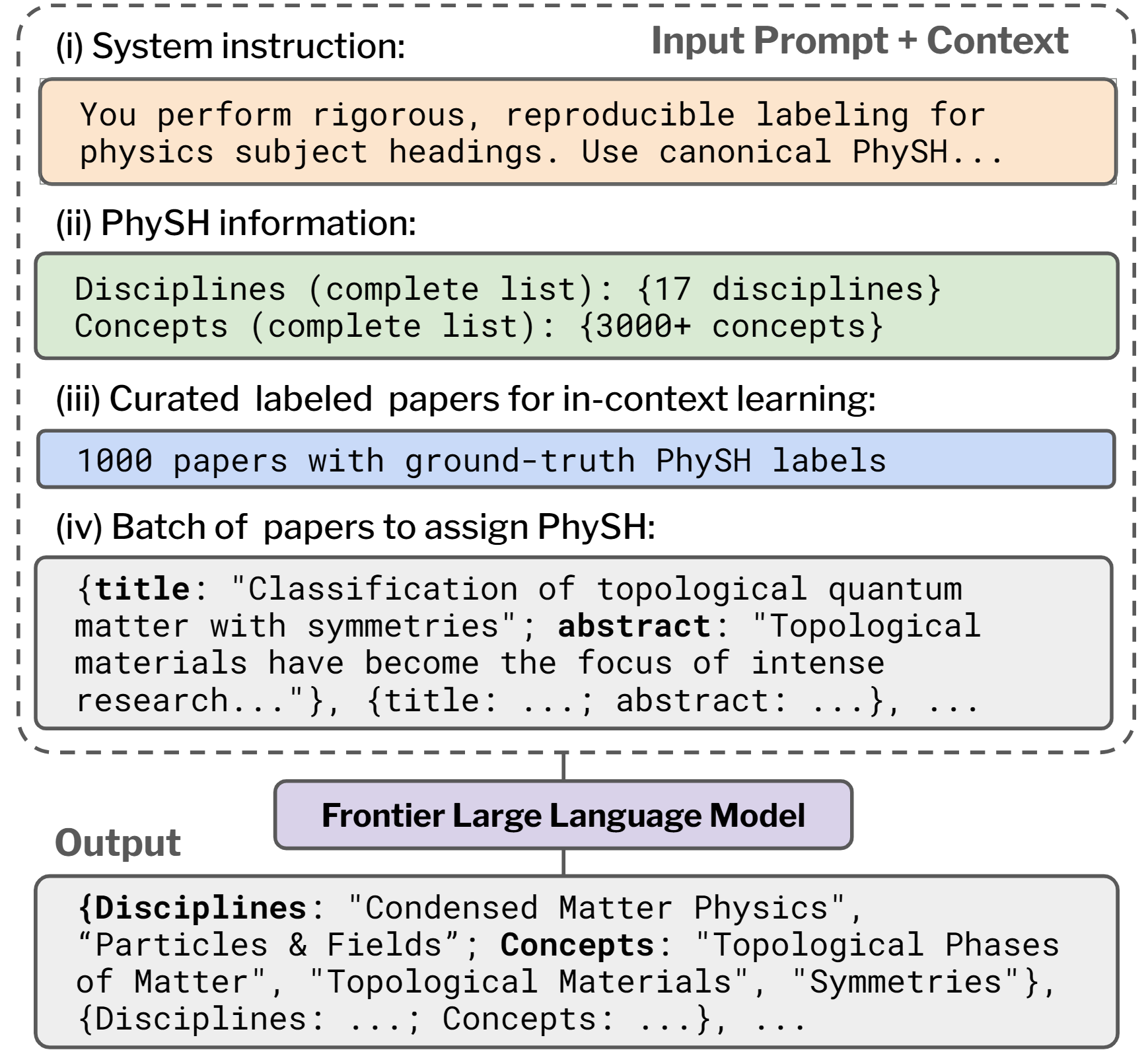}
    \caption{\textbf{LLM-based PhySH annotation pipeline.} The query contains the titles and abstracts of archival papers, combined with system instructions, the full PhySH taxonomy, and curated examples. The frontier LLM returns structured discipline and concept labels for each paper.}
    \label{fig:llm_pipeline}
\end{figure}

We generate PhySH labels for papers in batches using a frontier LLM, wherein the query contains four parts [Fig.~\ref{fig:llm_pipeline}]: (i) task instructions requiring rigorous, reproducible use of canonical PhySH names; (ii) the complete lists of valid disciplines and concepts; (iii) curated PhySH-labeled examples for in-context learning; and (iv) the titles and abstracts of the batch of papers to be labeled. The model output is constrained to structured discipline and concept lists. The full prompt template is given in Appendix A.

We benchmark candidate models using two complementary metrics on a representative set of post-2016 papers. Discipline accuracy is evaluated by sample-level F1 score \cite{Pereira2018MultilabelMeasures}. Concept accuracy is evaluated by semantic F1, in which near misses receive partial credit according to a PhySH concept-similarity matrix \cite{Chochlakis2025SemanticF1} (Appendix B). 

Figure~\ref{fig:metrics}(a) shows the tradeoff among accuracy, speed, and cost. The strongest models achieve similar performance, but differ substantially in throughput and cost per paper. We select Gemini 2.5 Flash Lite Preview \cite{GoogleGemini25FlashLiteAPI,GoogleGemini25FlashLiteBlog} because it provides high concept and discipline scores while enabling archive-scale labeling. We then optimize the number of examples and the batch size using a grid search [Fig.~\ref{fig:metrics}(b,c)]. On a separate 1,000-paper held-out benchmark, we achieve a concept semantic F1 score of 80.9\% and a discipline F1 score of 77.6\%, using a batch size of 50 and 1,000 few-shot examples per batch. To reduce variance from in-context example selection, we average all reported trajectories across five independent runs.

\begin{figure}[H]
    \centering
    \includegraphics[width=\linewidth]{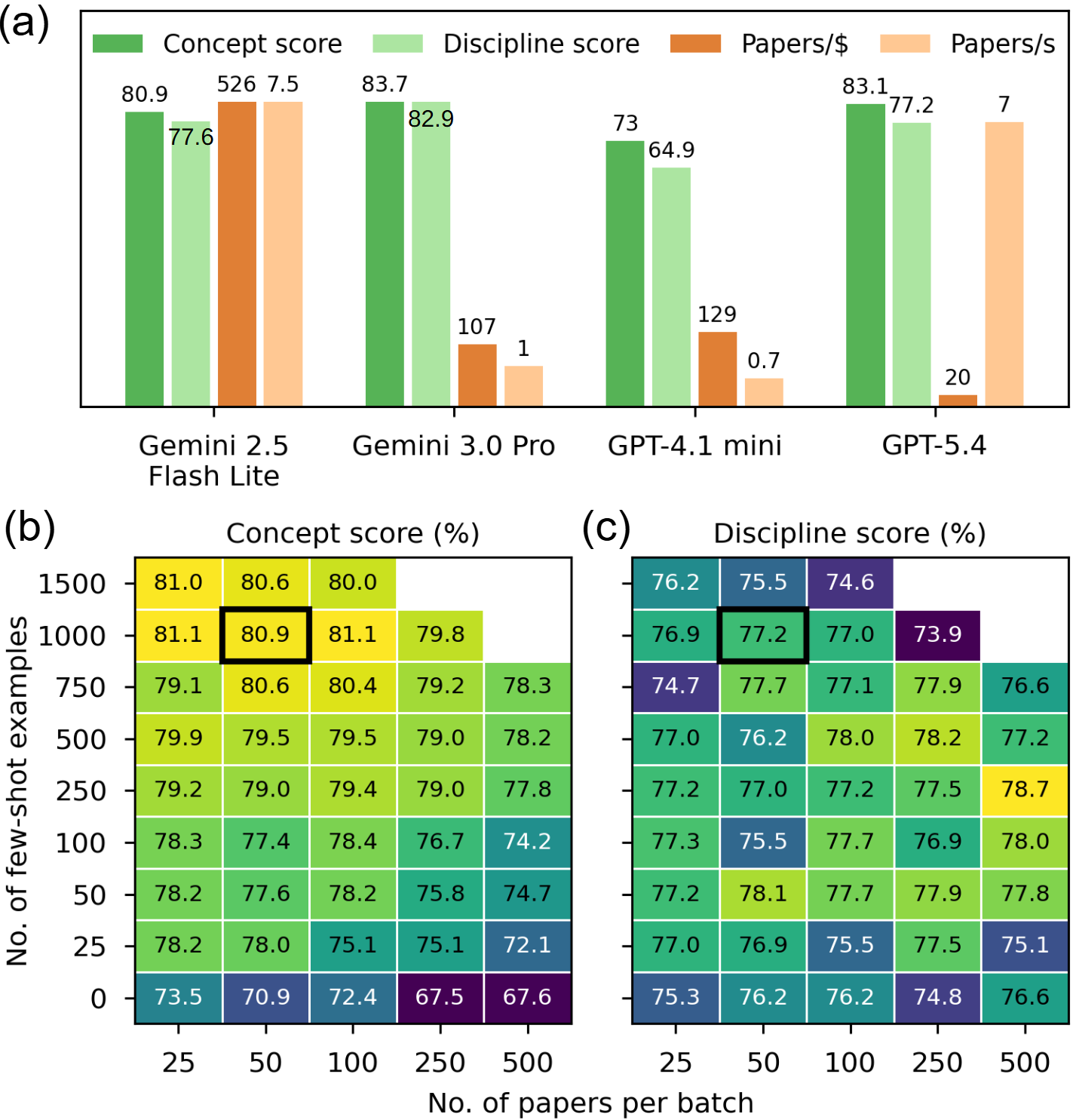}
    \caption{\textbf{Model selection and parameter optimization.}
    (a) Comparison of candidate LLMs by concept score, discipline score, cost, and speed. Cost estimates account for cached reuse of static prompt context where available.
    (b,c) Grid search over example count and batch size, showing (b) concept semantic F1 score and (c) discipline F1 score. The black outline marks the selected configuration.}
    \label{fig:metrics}
\end{figure}

\begin{figure*}[t]
    \centering
    \includegraphics[width=\linewidth]{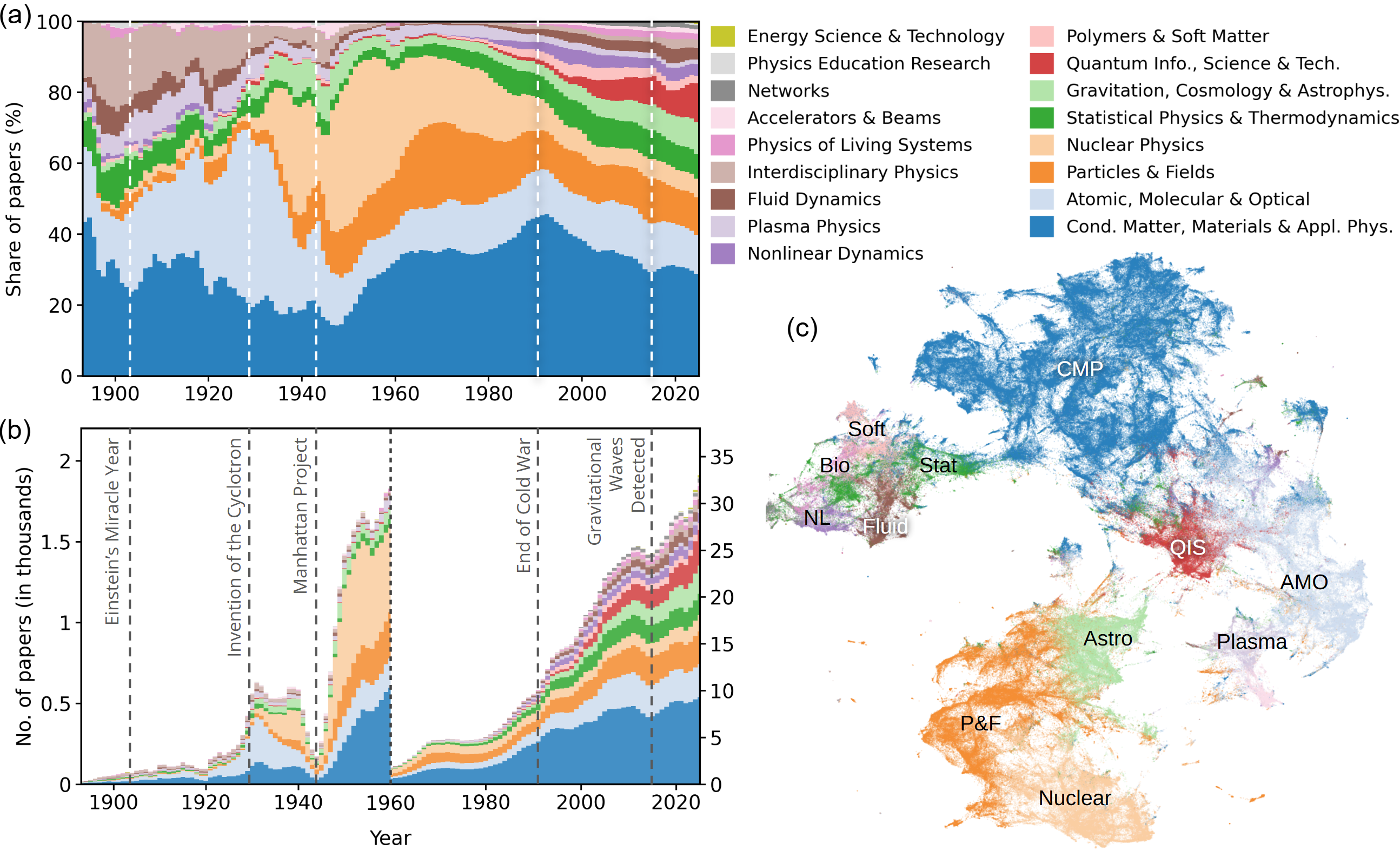}
    \caption{\textbf{Disciplinary evolution of 130 years of physics.}
    Share of papers (a) and raw numbers of papers (b) assigned to each PhySH discipline from 1893 to 2025, shown as three-year centered rolling averages. Dashed vertical lines mark selected historical milestones. The vertical scale is zoomed-in pre-1960 in (b) (dotted line). All plotted discipline counts and shares are weighted averages over the five runs. For papers assigned multiple disciplines, each assigned discipline receives equal fractional weight.
    (c) Two-dimensional UMAP projection of title and abstract embeddings, colored by assigned primary discipline.}
    \label{fig:disciplines}
\end{figure*}

 \begin{figure*}[t]
    \centering
    \includegraphics[width=\linewidth]{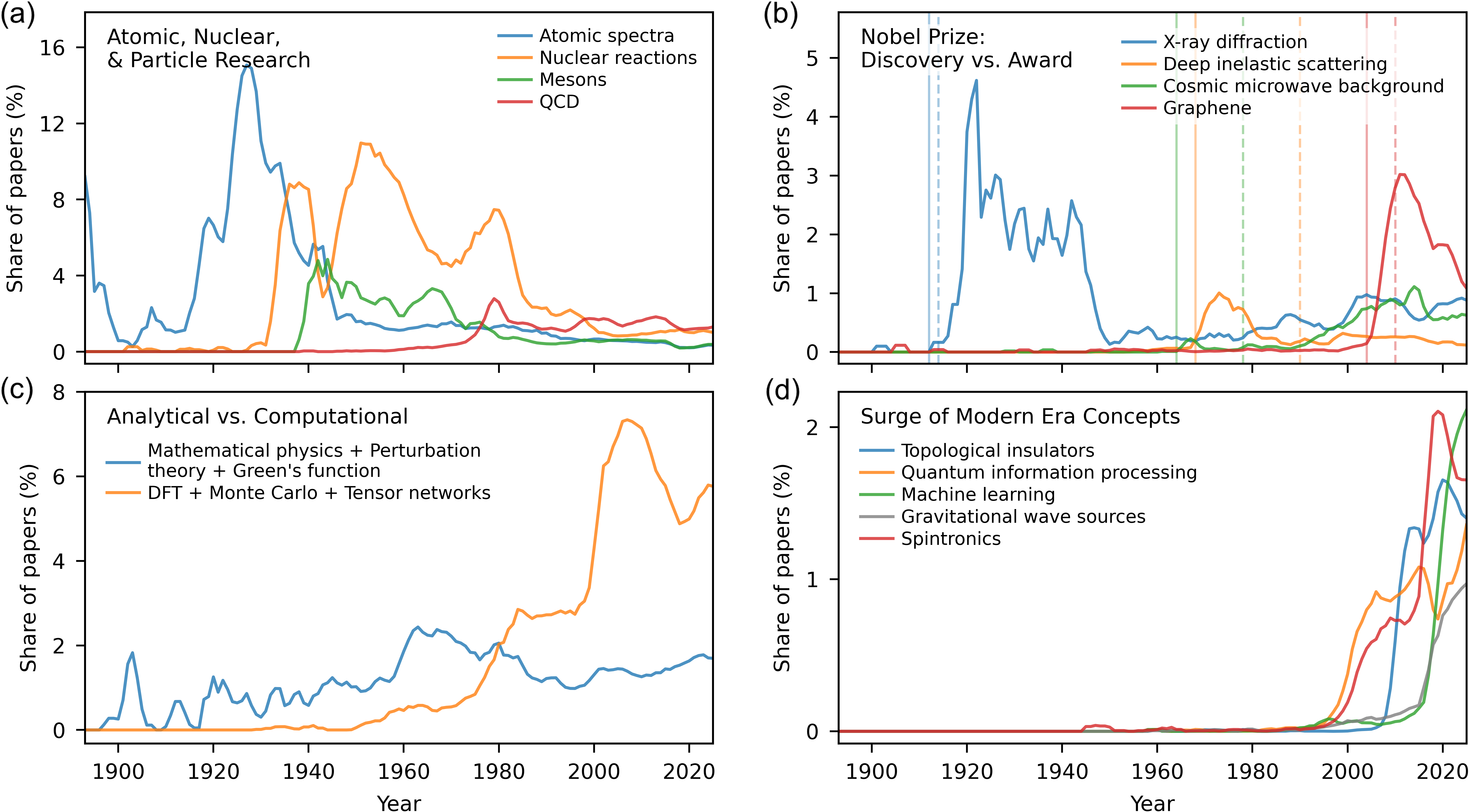}
    \caption{\textbf{Trajectories of selected concepts.}
    Share of papers tagged with selected PhySH concepts, shown as three-year centered rolling averages. All trajectories are averaged over five independent runs.
    (a) Sequential trajectories linking early atomic physics, mid-century nuclear and particle physics, and later fundamental field-theory descriptions.
    (b) Nobel-awarded concept lifecycles; solid vertical lines mark discoveries and dashed vertical lines mark prize years.
    (c) Comparison of traditionally analytical with increasingly compute-intensive frameworks. Aggregate curves show the summed shares of the listed concepts.
    (d) Rapid growth of recent concepts.}
    \label{fig:concepts}
\end{figure*}

The resulting subject map is a continuous 130-year record of physics under a shared vocabulary. Using it, we first visualize the evolution of the 17 broad disciplines, highlighting the impact of major historical events, before analyzing the trajectory of specific ideas, methods, and discoveries captured by the faceted PhySH concepts.  

\textit{Disciplinary Arcs---}The evolution of the 17 PhySH disciplines between 1893 and 2025 is shown in Fig.~\ref{fig:disciplines}(a,b). Condensed Matter, Materials \& Applied Physics (CMP) and Atomic, Molecular, \& Optical Physics (AMO) remain prominent throughout the record, though with distinct temporal profiles: AMO accounts for a large share of early twentieth-century research, especially in the 1920s and 1930s, while CMP becomes the largest discipline by the 1970s.

On the other hand, Nuclear Physics (NP) rises rapidly in the 1930s, consistent with the invention of the cyclotron \cite{LawrenceLivingston1932Cyclotron} and the discovery of nuclear fission \cite{HahnStrassmann1939Barium,MeitnerFrisch1939}. By the end of the decade, NP had become the dominant discipline in the archive, and its share continues to grow through the World War II era before declining after the 1960s and toward the end of the Cold War. Surprisingly, during the war itself, NP's share declines even as its strategic importance increases and the total number of publications plummets, potentially due to the redirection of nuclear research into classified programs \cite{QuistClassification,NPSManhattanProject}. Particles \& Fields (P\&F) follows a related but distinct postwar arc, consistent with the founding of CERN in 1954 \cite{CERNHistory} and the emergence of high-energy physics as a central route to fundamental investigation. P\&F retains a substantial share after the Cold War.

In the post-Cold War era, the disciplinary distribution becomes more balanced. Quantum Information, Science \& Technology (QIS) emerges following demonstrations of quantum-coherent circuits \cite{Martinis1985EnergyLevelQuantization,Devoret1985MacroscopicTunneling} and the development of Shor's algorithm \cite{Shor1994Algorithms}. Gravitation, Cosmology \& Astrophysics (GCA) gains prominence as well, with LIGO's first direct detection of gravitational waves marking a major modern milestone \cite{Abbott2016GravitationalWaves}. Polymers \& Soft Matter, Nonlinear Dynamics, Physics of Living Systems, and Networks all gain visible share, transitioning the archive from being shaped by a few dominant programs to one with a more distributed disciplinary structure.

This map can also be visualized in the semantic space with LLM-derived embeddings \cite{Cohan2020SPECTER,Singh2023SciRepEval}. We embed each paper's title and abstract using the Gemini Embedding model \cite{Lee2025GeminiEmbedding} and project the resulting vectors onto two dimensions [Fig.~\ref{fig:disciplines}(c)]. This visualization reveals coherent discipline-level regions and three broad superclusters: a quantum-matter and devices cluster containing much of CMP, AMO, and QIS; a complex-systems cluster containing Soft Matter, Biological Physics, Statistical Physics, Fluids, and Nonlinear Dynamics; and a fundamental-physics cluster containing NP, P\&F, and GCA.

\textit{Concept Trajectories---}Discipline-level histories reveal broad shifts in physics, but PhySH enables a finer view by resolving the concepts that track individual topics, ideas, materials, and methods. In Fig.~\ref{fig:concepts}, we plot the yearly share of papers with selected concepts, so that each trajectory measures a concept's relative prominence within the archive.

First, we trace a sequence from atomic spectroscopy through nuclear and meson physics to field-theoretic descriptions [Fig.~\ref{fig:concepts}(a)]. The early prominence of Atomic spectra reflects the central role of spectroscopy in establishing the empirical foundations of quantum mechanics \cite{Bohr1913,Bohr1918LineSpectra}. As accelerators developed and attention shifted from electronic structure toward the nucleus, Nuclear reactions rose sharply \cite{MeitnerFrisch1939,DOEManhattanProject}. The subsequent rise of Mesons captures the postwar ``particle zoo'' era \cite{BrownDresdenHoddeson1988,Harlander2023ParticleEra}, which was later consolidated and unified by the development of Quantum chromodynamics (QCD) \cite{GellMann1964,GrossWilczek1973,Politzer1973,Wilczek2005QCD}.

We next examine four Nobel-awarded concepts as time-stamped case studies for influential ideas in physics, highlighting distinct trajectories across discovery, prize recognition, and subsequent activity [Fig.~\ref{fig:concepts}(b)]. X-ray diffraction was recognized almost immediately after its 1912 discovery, before its broad adoption as a structural probe \cite{FriedrichKnippingLaue1912,Nobel1914Laue}. It remained intensely researched until the early 1940s, after which it sharply declined in APS journals, likely reflecting maturation and migration into more specialized literatures \cite{Bragg1962GrowingPower,Wyckoff1962XrayUSA}. Deep inelastic scattering shows the opposite timing: its sharp rise followed the initial experiments in the late 1960s, whereas Nobel recognition came only in 1990, after the concept had passed its peak as a frontier physics field \cite{Bloom1969DIS,Breidenbach1969DIS,Nobel1990DIS}. Cosmic microwave background follows a delayed-growth trajectory: its 1965 identification initially saw limited activity, but later catalyzed sustained growth, notably with the 1978 Nobel recognition and the commissioning of major new instruments in the 1980s--1990s \cite{PenziasWilson1965,Dicke1965,Nobel1978CMB,Smoot1992}. Finally, Graphene shows a compressed modern lifecycle, rising rapidly after its 2004 isolation and peaking near its 2010 Nobel recognition \cite{Novoselov2004,Nobel2010Graphene}.

Figure~\ref{fig:concepts}(c) captures the shift from primarily analytical to increasingly computational approaches in physics. We compare traditionally analytical topics, represented by Mathematical physics, Perturbation theory, and Green's functions, with computational ones, represented by Density functional theory, Monte Carlo methods, and Tensor networks. The latter aggregate grows strongly in the late twentieth and early twenty-first centuries, consistent with the development and widespread adoption of these computational frameworks \cite{Metropolis1953,HohenbergKohn1964,KohnSham1965,White1992,Jones2015DFT,Schollwock2011,Orus2019TensorNetworks} and the co-evolution of computing hardware \cite{DongarraKeyes2024}.

Finally, Fig.~\ref{fig:concepts}(d) illustrates the rapid growth of recent concepts, including Topological insulators, Quantum information processing, Machine learning, Gravitational wave sources, and Spintronics. These curves rise on relatively shorter timescales than many earlier topics, reflecting a contemporary research environment in which new ideas, methods, materials, and data are shared and disseminated rapidly across the globe.

\textit{Discussion and Outlook---}We have shown that LLMs can retrospectively assign PhySH labels to the historical APS corpus, producing a unified and searchable taxonomy that spans more than a century of physics research. 
Nonetheless, because labels are inferred from titles and abstracts rather than full texts, and are benchmarked against author-assigned PhySH labels that themselves contain human variability, the resulting trajectories should be interpreted as subject-level trends rather than definitive classifications of individual papers.

This approach could be readily extended beyond APS journals and English-language papers to provide a broader view of the evolution of physics. Such extensions would enable quantitative comparisons of research trajectories across journals, languages, and regions with subject-level resolution.

Historical PACS codes could also provide a complementary prior for retrospective classification for the many journals where PACS was used, although exploiting it fully would require reliable extraction from full-text PDFs \cite{PACSHome,APSNews2016PhySH,PRL2016EditorialReminders}. Subject-level labels could also complement citation-dynamics studies \cite{Medo2011TemporalEffects,Golosovsky2012CitationNetwork,Wang2013ScientificImpact} by enabling analyses along explicit topical axes.

More broadly, LLMs may support the continued curation of the PhySH taxonomy by identifying redundant concepts, suggesting refinements to hierarchical structure, and tracking changes in scientific terminology over time. They could also assist authors and editors by suggesting PhySH labels for new submissions, reducing inconsistencies at the point of entry and helping maintain a coherent, searchable record as new research areas emerge.

\textit{Acknowledgments---}The authors thank APS for access to the dataset and APS Chief Information Officer Mark Doyle for helpful correspondence. E.Y.M. acknowledges support from a Google Research Scholarship.

\textit{Data availability---}The data and code used to generate the figures in this Letter are available in \cite{NguyenPandeyLiMa2026APSPhySH}. The released data include the DOIs and generated PhySH discipline and concept labels for all five runs. Additional metadata, including titles and abstracts, are available from APS.

\bibliographystyle{apsrev4-2}
\bibliography{references}

\clearpage
\appendix

\title{End Matter}
\endmattertitletrue
\maketitle

\setcounter{equation}{0}
\renewcommand{\theequation}{A\arabic{equation}}

\noindent\textit{Appendix A: Prompt Template and Labeling Protocol---}This appendix gives the prompt template used for PhySH labeling of APS papers from titles and abstracts. For papers without abstracts, the abstract field was passed as an empty string and labeling used the title alone.

Production labeling used Gemini 2.5 Flash Lite Preview with the selected configuration in Fig.~\ref{fig:metrics}(b,c): 1,000 in-context examples and batches of 50 papers per query. Each query contained task instructions, the complete allowed PhySH discipline and concept vocabularies, curated PhySH-labeled examples, and the titles and abstracts of the papers to be labeled. The model was required to return structured discipline and concept lists. Returned labels were validated against the allowed PhySH vocabularies, and malformed outputs or non-vocabulary labels were retried until a valid structured response was obtained. The 1,000-paper held-out benchmark was disjoint from all in-context examples used for model selection and production labeling. Other model settings were fixed across all production runs.

\begin{quote}
\small
You perform rigorous, reproducible labeling for physics subject headings.
Use canonical PhySH names; normalize common synonyms to the canonical form.
Identify the core problem and method from title and abstract before labeling.
Prefer clarity over verbosity; minimize non-essential wording.
Resolve ambiguity by preferring labels supported by explicit keywords.
When a method is emphasized, include an appropriate Techniques concept.
Resolve conflicts by deferring to precise technical usage over colloquial usage.

Prefer a single discipline when one field clearly dominates the core contribution.
Assign a second discipline when the paper makes a distinct, independently sufficient contribution to that field; do not under-assign if two fields are central.
If the paper develops or validates a method that is itself a central contribution, and not merely applied, include the method's discipline as a second discipline.
Rarely, assigning three disciplines is permitted if the paper fundamentally bridges three distinct fields with equal weight.
If three fields are equally central and independently justified, include all three; do not collapse to two.
The threshold for assigning a second or third discipline is high: the paper must advance the state of the art in each field.

When a paper spans multiple disciplines, verify that each assigned discipline is essential to the paper's core contribution.
If one is merely a context or tool, exclude the discipline, but retain specific concepts that accurately describe that method or tool.
Prioritize the single best-fitting discipline when evidence is insufficient for multiple fields.
If the paper applies established methods from one field to a problem in another, assign only the discipline of the problem being solved.
However, if this filtering results in zero disciplines, assign the single most relevant discipline representing the paper's general context.

Critical instruction: list concepts in order from broadest to most specific.
Start the labeling process with the major field, then the subfield, then the specific topic.
For example, use an ordering such as ``Semiconductors'' then ``Quantum dots.''
Explicitly including broad terms first ensures that they are not omitted.
Make sure the broad terms are present in the final JSON list.

Critical instruction: balance specificity and frequency.
First, assign a Property concept only if the paper investigates the nature or mechanism of that property.
If the property is only a known feature of the system, tag the system rather than the property.
Second, include broad Physical Systems labels for every specific system mentioned.
For example, if ``Graphene'' is assigned, then ``2-dimensional systems'' must also be assigned.
Third, for Techniques concepts, prefer standard, established method names, such as ``Monte Carlo methods.''
Fourth, for Research Areas, ensure that the broad research area is present when supported by the title and abstract, such as ``Electronic structure.''

Critical instruction: distinguish concepts from disciplines.
Do not simply repeat the discipline name as a concept.
In particular, ``Condensed Matter, Materials \& Applied Physics,'' ``Particles \& Fields,'' ``Atomic, Molecular \& Optical,'' and ``Gravitation, Cosmology \& Astrophysics'' are disciplines, not valid concepts.
Do not assign them as concepts.
Concepts must describe specific topics, systems, properties, techniques, or research areas within the field.

Critical instruction: apply the discipline assignment rules conservatively.
``Quantum Information'' and ``Statistical Physics'' are often used as tools in Condensed Matter.
If the paper makes a substantive contribution to Quantum Information or Statistical Physics methods or theory, rather than merely applying them, assign the corresponding discipline.
If the paper applies standard Quantum Information or Statistical Physics tools to a Condensed Matter system, the discipline is only ``Condensed Matter, Materials \& Applied Physics.''

Represent both main phenomena and relevant theoretical or experimental frameworks when clearly described.
Normalize morphological variants to the canonical PhySH label.
Do not return empty discipline or concept lists; for each paper, return at least one discipline and at least one concept.
Avoid patterns that would systematically inflate rare concepts without strong evidence.
Do not invent new labels or paraphrase existing ones.
Adhere to the allowed labels without deviation.

Examples (gold):
\[
\begin{array}{l}
\texttt{[}\\
\quad \texttt{\{"text\_input":"<title + abstract>",}\\
\quad \texttt{"disciplines":[...],}\\
\quad \texttt{"concepts":[ ...]\},}\\
\quad \texttt{ ... batch of 1,000  labeled examples}\\
\texttt{]}
\end{array}
\]

Use only these valid disciplines:
\[
\{\texttt{complete PhySH discipline list}\}
\]

Use only these valid concepts:
\[
\{\texttt{complete PhySH concept list}\}
\]

User:
\[
\begin{array}{l}
\texttt{[}\\
\quad \texttt{\{"id":0,"text\_input":"..."\},}\\
\quad \texttt{\{"id":1,"text\_input":"..."\},}\\
\quad \texttt{ ... batch of 50  papers to be labeled}\\
\texttt{]}
\end{array}
\]
\end{quote}

\setcounter{equation}{0}
\renewcommand{\theequation}{B\arabic{equation}}

\noindent\textit{Appendix B: Evaluation Metrics---}We evaluate model performance on a held-out benchmark using two label-set metrics: an exact sample F1 score for disciplines and a semantic sample F1 score for concepts.

For \emph{disciplines}, predictions are scored by exact agreement with the author-assigned PhySH discipline labels. For paper $i$, let $D_i$ be the predicted discipline set and $G_i$ the ground-truth set. We define discipline precision and recall as
\begin{equation}
P_i^{(d)}
=
\frac{|D_i \cap G_i|}{|D_i|},
\end{equation}
and
\begin{equation}
R_i^{(d)}
=
\frac{|D_i \cap G_i|}{|G_i|}.
\end{equation}
The sample-level discipline F1 score is their harmonic mean,
\begin{equation}
F1_i^{(d)}
=
\frac{2P_i^{(d)}R_i^{(d)}}{P_i^{(d)}+R_i^{(d)}}
=
\frac{2|D_i \cap G_i|}{|D_i|+|G_i|},
\end{equation}
and the reported discipline score is the mean over all $N$ held-out papers,
\begin{equation}
F1^{(d)}
=
\frac{1}{N}\sum_{i=1}^{N}F1_i^{(d)}.
\end{equation}

For \emph{concepts}, exact matching is overly strict because nearby PhySH concepts can describe closely related scientific content. We therefore use the sample Semantic F1 framework of Ref.~\cite{Chochlakis2025SemanticF1}. Let $\mathcal{L}$ denote the full PhySH concept vocabulary, and let
$S\in[0,1]^{|\mathcal{L}|\times|\mathcal{L}|}$ be a concept-similarity matrix, where $S_{ab}$ measures the semantic similarity between concepts $a,b\in\mathcal{L}$. For any two nonempty concept sets $A,B\subseteq\mathcal{L}$, define
\begin{equation}
\mathrm{BestMatch}(A,B;S)
=
\frac{1}{|A|}
\sum_{a\in A}\max_{b\in B}S_{ab}.
\end{equation}
This quantity measures how well each concept in $A$ is covered by its closest semantic match in $B$.

For paper $i$, let $C_i\subseteq\mathcal{L}$ be the predicted concept set and $T_i\subseteq\mathcal{L}$ the ground-truth concept set. Semantic precision and recall are
\begin{equation}
P_i^{(c,\mathrm{sem})}
=
\mathrm{BestMatch}(C_i,T_i;S),
\end{equation}
and
\begin{equation}
R_i^{(c,\mathrm{sem})}
=
\mathrm{BestMatch}(T_i,C_i;S).
\end{equation}

The concept semantic F1 score for paper $i$ is

\begin{equation}
F1_i^{(c,\mathrm{sem})}
=
\frac{
2P_i^{(c,\mathrm{sem})}R_i^{(c,\mathrm{sem})}
}{
P_i^{(c,\mathrm{sem})}+R_i^{(c,\mathrm{sem})}
},
\end{equation}

and the reported concept score is the mean over all $N$ held-out papers,

\begin{equation}
F1^{(c,\mathrm{sem})}
=
\frac{1}{N}\sum_{i=1}^{N}F1_i^{(c,\mathrm{sem})}.
\end{equation}

When $S$ is the identity matrix, this expression reduces to ordinary sample F1. We construct $S$ from Gemini embeddings of concept strings containing the PhySH concept's name, discipline, parent concepts, and child concepts. Let $\mathbf{e}_a$ denote the embedding of concept $a\in\mathcal{L}$. We first compute normalized cosine similarities,
\begin{equation}
\tilde{S}_{ab}
=
\frac{1+\cos(\mathbf{e}_a,\mathbf{e}_b)}{2},
\end{equation}
and then apply a single global affine rescaling,
\begin{equation}
S_{ab}
=
\frac{\tilde{S}_{ab}-m}{1-m},
\qquad
m=\min_{\substack{u,v\in\mathcal{L}\\u\neq v}}\tilde{S}_{uv}.
\end{equation}
Here the minimum is taken over all distinct pairs of PhySH concepts in the vocabulary.

\end{document}